\documentclass[aps,prl,twocolumn,superscriptaddress]{revtex4}
\usepackage{epsf,graphicx}
\usepackage{amssymb}
\usepackage{amsmath}
\usepackage{latexsym,bm,array,amsfonts,multirow}
\usepackage{color}
\usepackage{ulem}
\begin{document}

\title{Possible Fano effect and suppression of Andreev reflection in La$_3$Ni$_2$O$_7$}
\author{Yi-feng Yang}
\email[]{yifeng@iphy.ac.cn}
\affiliation{Beijing National Laboratory for Condensed Matter Physics and Institute of
Physics, Chinese Academy of Sciences, Beijing 100190, China}
\affiliation{University of Chinese Academy of Sciences, Beijing 100049, China}
\affiliation{Songshan Lake Materials Laboratory, Dongguan, Guangdong 523808, China}
\date{\today}

\begin{abstract}
The recently-discovered high-temperature superconductor La$_3$Ni$_2$O$_7$ under high pressure has stimulated intensive debates. Key controversies concern interlayer versus intralayer pairing scenarios and if the hybridization plays a key role in establishing the superconductivity. But experimental clarification is difficult due to the limitation of employing state-of-the-art techniques under high pressure. Here we propose that quasiparticle tunneling and Andreev reflection may provide a feasible way to distinguish different pairing scenarios. We predict that an asymmetric Fano line shape may be induced by the hybridization between the $d_{x^2-y^2}$ metallic bands and the strongly renormalized flat $d_{z^2}$ quasiparticle bands. In the superconducting state, we show that the Andreev reflection should be greatly suppressed for interlayer pairing superconductivity with a small interlayer hopping. We propose future experiments to examine these predictions and help clarify the fundamental physics of superconducting La$_3$Ni$_2$O$_7$ and other multilayer nickelate superconductors.
\end{abstract}
\maketitle

The recent discovery of high-temperature superconductivity in the bilayer nickelate La$_3$Ni$_2$O$_7$ under high pressure has stimulated intensive theoretical debates concerning its electronic band structures and potential pairing mechanism \cite{MWang2023Nature, JGCheng2023, HQYuan2024}. Although experiments at ambient pressure have helped to clarified many important issues \cite{HHWen2023, XJZhou2024, LShu2024, ZChen2024, LYang2024, MWang2024, DLFeng2024, XChen2024, ZXShen2024, LYang2024b, TXiang2024review}, direct probes of its superconducting properties are very challenging due to the high pressure. One of the fundamental issues involves the nature of its pairing mechanism. While many propose interlayer pairing mediated by the interlayer superexchange interaction of $d_{z^2}$ orbitals \cite{GMZhang2023,QQin2023,QHWang2023,YFYang2023,FYang2023,JPHu2023,WWu2023,Zhang2023d,GSu2023,Kuroki2024,CJWu2024,ZYLu2024,Heier2023,WLi2024,DXYao2023tJ,FWang2024,Wang2024arxiv}, some argue that the superconductivity is similar to that in cuprates \cite{TXiang2023,KJiang2024,WKu2024}. Among all interlayer pairing scenarios, debates also exist concerning the role of the Hund's rule coupling and the hybridization between the $d_{x^2-y^2}$ and $d_{z^2}$ orbitals \cite{YFYang2023,ZYLu2024,CJWu2024,GSu2023b}. In the two-component scenario, it is the hybridization that helps mobilize the local $d_{z^2}$ interlayer pairs and induces the long-range phase coherence \cite{QQin2023,YFYang2023}. While in the Hund scenario, the hybridization and the self-doping to the $d_{z^2}$ orbitals are both harmful, and it is the Hund's rule coupling that transfers the interlayer superexchange interaction between localized $d_{z^2}$ orbitals to the itinerant $d_{x^2-y^2}$ orbitals to form interlayer Cooper pairing \cite{CJWu2024}. It is therefore important to distinguish the intralayer and interlayer pairing mechanisms and provide experimental evidences for the presence of the hybridization.

However, high pressure has prevented the application of many important techniques. Among the very few exceptions, quasiparticle tunneling and Andreev reflection may be a feasible way to provide some useful microscopic information on the electronic and superconducting properties. In this work, we argue that they might indeed exhibit distinct properties and help distinguish different pairing scenarios in superconducting La$_3$Ni$_2$O$_7$. In particular, we show the possibility of Fano effect in the quasiparticle tunneling spectra if hybridization indeed exists between the broad $d_{x^2-y^2}$ metallic bands and the strongly renormalized flat $d_{z^2}$ quasiparticle bands, and that, in contrast to intralayer pairing superconductivity, the Andreev reflection may be greatly suppressed for interlayer pairing with small interlayer hopping. We suggest future experiments to verify these predictions.

We start with the following Hamiltonian containing three parts:
\begin{equation}
H=H_{\rm s} + H_{\rm m} + H_{\rm c}, 
\end{equation}
where $H_{\rm s}$, $H_{\rm m}$, and $H_{\rm c}$ describe the nickelate superconductor, the normal metal lead, and their coupling term, respectively. The La$_3$Ni$_2$O$_7$ Hamiltonian takes an effective two-orbital bilayer form \cite{QQin2023,YFYang2023,Wang2024arxiv}:
\begin{equation}
\begin{split}
H_{\rm s}=&\sum_{la\mathbf{k}s}\epsilon^a_\mathbf{k}d_{la\mathbf{k}s}^{\dagger}d_{la\mathbf{k}s}-\sum_{l\mathbf{k}s}V_{\mathbf{k}}\left(d_{l1\mathbf{k}s}^{\dagger}d_{l2\mathbf{k}s}+h.c.\right)\\
-&t_\perp\sum_{\mathbf{k}s} \left(d_{11\mathbf{k}s}^\dagger d_{21\mathbf{k}s}+h.c.\right)+J\sum_{i}\bm{S}_{1i}\cdot\bm{S}_{2i},
\end{split}
\end{equation}
where $d^\dagger_{la\mathbf{k}\sigma}$ creates one $d_{z^2}$ ($a=1$) or $d_{x^2-y^2}$ ($a=2$) electron of spin $s$ and momentum $\mathbf{k}$ on $l$-th layer ($l=1,2$) $\bm{S}_{li}=\frac12\sum_{ss'}d_{l1is}^{\dagger}\bm{\sigma}_{ss'}d_{l1is'}$ is the $d_{z^2}$ spin density operator on site $i$ of $l$-th layer with $\boldsymbol{\sigma}$ being the Pauli matrices, $\epsilon^a_{\mathbf{k}}$ gives the dispersion of $a$-th orbital within each layer, $V_{\mathbf{k}}$ is the intralayer nearest-neighbor hybridization between two orbitals, $t_\perp$ is the interlayer hopping of $d_{z^2}$ quasiparticles, and $J$ is their interlayer superexchange interaction mediated by the apical oxygens. 

The above effective $t$-$V$-$J$ model is similar to the effective $t$-$J$ model for cuprate superconductors and may be derived by projecting out the double occupancy of the $d_{z^2}$ orbitals from the bilayer two-orbital Hubbard model. As in cuprates, oxygen degrees of freedom are integrated out. Whether or not a more complicated model containing all orbitals is necessary in some circumstances requires future experimental scrutiny. First-principles calculations give $\epsilon^a_{\mathbf{k}}=-2t_a(\cos\mathbf{k}_x+\cos\mathbf{k}_y)+4t'_a\cos\mathbf{k}_x\cos\mathbf{k}_y+\epsilon_0^a$ and $V_\mathbf{k}=-2V(\cos\mathbf{k}_x-\cos\mathbf{k}_y)$, where $t_1=0.11$, $t'_1=-0.017$, $\epsilon_0^1=0.409$, $t_2=0.483$, $t'_2=0.069$, $\epsilon_0^2=0.776$, $V=0.239$, and $t_\perp=0.635$ \cite{DXYao2023}. For simplicity, we drop the energy unit eV throughout this paper unless noted. Other small parameters are ignored for simplicity and their effects may be partly included by renormalizing the above tight-binding parameters \cite{YFYang2023,YYCao2024}. The intralayer superexchange interaction is also ignored to focus only on the interlayer pairing since it is ten times smaller than the interlayer superexchange $J$ as estimated lately in the resonant inelastic X-ray scattering (RIXS) and inelastic neutron scattering (INS) measurements \cite{MWang2024, DLFeng2024}. The lead and their coupling term can be described as 
\begin{equation}
H_{\rm m}=\sum_{\mathbf{k}s}\xi_\mathbf{k}c_{\mathbf{k}s}^\dagger c_{\mathbf{k}s},\ \ \ H_{\rm c}=\sum_{a\mathbf{kp}s}t^a_{\mathbf{kp}} c_{\mathbf{k}s}^\dagger d_{1a\mathbf{p}s},
\end{equation}
where $\xi_\mathbf{k}$ is the dispersion of the noninteracting electrons in the normal metal lead and $t^a_\mathbf{k}$ is the tunneling matrix into the $a$-th orbital. For simplicity, we only consider explicitly the momenta $\mathbf{k}$ in the plane.

\begin{figure}[t]
\centering
\includegraphics[width=0.7\linewidth]{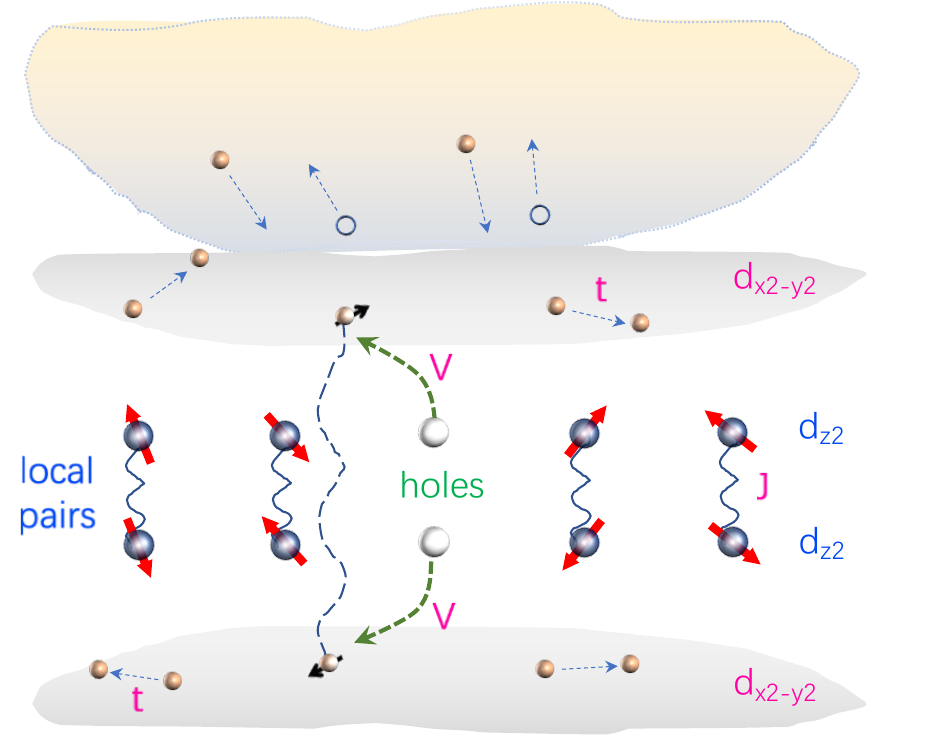}
\caption{Illustration of the the normal metal/interlayer pairing superconductor junction and the effective $t$-$V$-$J$ model for the bilayer nickelate,  where $t$ is the hopping parameter of the metallic band, $J$ is the interlayer superexchange interaction of the local pairing orbital, and $V$ is their hybridization. The electrons may tunnel into both the broad $d_{x^2-y^2}$ bands and the relatively narrow $d_{z^2}$ bands in the top layer. The Fano interference arises because of the hybridization of the two bands. The Andreev reflection may be reduced when $t_\perp\rightarrow 0$ because of the interlayer pairing mechanism.}
\label{fig1}
\end{figure}

Figure \ref{fig1} gives a schematic illustration of the whole setup. The quasiparticle tunneling and Andreev reflection can be studied using the Keldysh Green's function technique \cite{Haug1998}. Applying a bias voltage $V_b$ on the lead induces a finite electric current \cite{Meir1992PRL,Yang2001PRB,Dai2024PRB}:
\begin{equation}
\begin{split}
I=&-e\frac{dN_c}{dt}=-i\frac{e}{\hbar}\sum_{a\mathbf{kq}s}t^a_{\mathbf{kq}}\left\langle d_{1a\mathbf{q}s}^\dagger c_{\mathbf{k}s}\right \rangle + c.c.\\
=&\frac{e}{h}\int d\omega \left[f(\omega-eV_b)-f(\omega)\right]T(\omega),
\end{split}
\end{equation}
where $N_c=\sum_{\mathbf{k}s}c_{\mathbf{k}s}^\dagger c_{\mathbf{k}s}$ is the total number of electrons in the lead, $f(\omega)$ is the Fermi distribution function, and we have the scattering matrix:
\begin{equation}
T(\omega)=\sum_{aa'a''\mathbf{kpq}s}\Gamma^0_{\mathbf{k}}\Gamma^{aa'}_{\mathbf{qp}}\mathcal{G}^R_{a'\mathbf{p}s,a''\mathbf{k}s}(\omega)\mathcal{G}^A_{a''\mathbf{k}s,a\mathbf{q}s}(\omega),
\end{equation}
where $\Gamma^0_\mathbf{k}$ measures the quasiparticle line width in the top NiO layer, $\Gamma^{aa'}_{\mathbf{qp}}=2\pi \sum_\mathbf{k'}t^a_{\mathbf{k'q}}  t^{a'}_{\mathbf{pk'}} \rho^c_{\mathbf{k'}s}(\omega)$ with $\rho^c_{\mathbf{k'}s}(\omega)$ being the spectral function of the lead electrons, and $\mathcal{G}^{R/A}$ denote the retarded/advanced Green's functions of the top layer ($l=1$) in the Nambu representation $\psi_\mathbf{k}^\dagger=\left(d_{11\mathbf{k}\uparrow}^\dagger, d_{12\mathbf{k}\uparrow}^\dagger, d_{11\bar{\mathbf{k}}\downarrow}, d_{12\bar{\mathbf{k}}\downarrow}\right)$. The total scattering can thus be split into two terms, $T(\omega)=T^Q(\omega)+T^A(\omega)$, where $T^{Q}$ and $T^{A}$ describe the quasiparticle tunneling and Andreev reflection contributions, respectively. We have
\begin{equation}
\begin{split}
T^Q(\omega)&=\sum_{aa'a''\mathbf{kpq}s}\Gamma^0_{\mathbf{k}}\Gamma^{aa'}_{\mathbf{qp}}G^R_{a'\mathbf{p}s,a''\mathbf{k}s}(\omega)G^A_{a''\mathbf{k}s,a\mathbf{q}s}(\omega),\\
T^A(\omega)&=\sum_{aa'a''\mathbf{kpq}s}\Gamma^0_{\bar{\mathbf{k}}}\Gamma^{aa'}_{\mathbf{qp}}F^R_{a'\mathbf{p}s,a''\bar{\mathbf{k}}\bar{s}}(\omega)F^{\dagger,A}_{a''\bar{\mathbf{k}}\bar{s},a\mathbf{q}s}(\omega),
\end{split}\label{TQTA}
\end{equation}
where $G^{R/A}$ and $F^{R/A}$ are the normal and anomalous components of the Green's function $\mathcal{G}^{R/A}$ in the Nambu representation, respectively. In the following, we will discuss $T^Q$ and $T^A$ separately.

In the above formalism, only the Green's functions of the top layer ($l=1$) are explicitly involved. To calculate them, we employ the mean-field approximation and decouple the superexchange term, $J\bm{S}_{1i}\cdot\bm{S}_{2i}\rightarrow \sqrt{2}\left(\Delta^*\Phi_{i}+\bar{\Phi}_i\Delta\right)$, where $\Phi_{i}=\frac{1}{\sqrt{2}}\left(d_{11i\downarrow}d_{21i\uparrow}-d_{11i\uparrow}d_{21i\downarrow}\right)$ represents the local interlayer $d_{z^2}$ pairing at site $i$ and $\Delta$ is the corresponding static uniform pairing field. Note that interlayer pairing of $d_{x^2-y^2}$ electrons will be induced through the hybridization as discussed in Refs. \cite{QQin2023,YFYang2023}. The mean-field approximation ignores the correlation effects as well as the spatial and temporal fluctuations of the superconductivity, so that all energy scales may be strongly renormalized and altered in real experiment, in particularly near the Fermi energy. Nevertheless, it still provides a reasonable starting point for studying qualitative features of the quasiparticle tunneling and Andreev reflection. Under the mean-field approximation, the Hamiltonian takes a bilinear form which can be solved straightforwardly in the momentum space. However, the coupling to the lead may introduce momentum scattering and complicate the calculations. To simplify the calculations, we ignore the off-diagonal contribution in $\mathcal{T}^{Q/A}$ by replacing $G_{\mathbf{pk}}\rightarrow G_{\mathbf{k}}\delta_\mathbf{pk}$ and $F_{\mathbf{p\bar{k}}}\rightarrow F_{\mathbf{k}}\delta_\mathbf{pk}$ or equivalently $\Gamma^{aa'}_{\mathbf{qp}}\rightarrow \Gamma^{aa'}_{\mathbf{k}}\delta_\mathbf{pk}\delta_\mathbf{qk}$ in Eq. (\ref{TQTA}). This may have some quantitative influence but does not affect our major conclusions that only depend on the presence or absence of the hybridization and the interlayer coupling. 

The Green's functions  for the top layer are then obtained by integrating out the electron degrees of freedom of the lead and the bottom layer ($l=2$). Under the Nambu basis $\psi_\mathbf{k}$, we derive in the matrix form,
\begin{equation}
(\mathcal{G}_\mathbf{k}^{R})^{-1}=g_\mathbf{k}^{-1} I-\mathcal{H}_\mathbf{k}^0-\Sigma_\mathbf{k}-M\left[g_\mathbf{k}^{-1} I-\mathcal{H}_\mathbf{k}^0\right]^{-1}M^\dagger,
\end{equation}
where $g_\mathbf{k}^{-1}=\omega+i\Gamma^0_\mathbf{k}/2$, $I$ is the $4\times4$ unit matrix, and
\begin{equation}
\begin{split}
&\mathcal{H}^0_\mathbf{k}=\left(\begin{array}{cccc}
\epsilon_\mathbf{k}^1&-V_\mathbf{k} &0&0\\
-V_\mathbf{k} &\epsilon_\mathbf{k}^2&0&0\\
0&0&-\epsilon_{\bar{\mathbf{k}}}^1&V_{\bar{\mathbf{k}}}\\
0&0 &V_{\bar{\mathbf{k}}}&-\epsilon_{\bar{\mathbf{k}}}^2\\
\end{array}\right),
M=\left(\begin{array}{cccc}
-t_\perp&0&\Delta&0\\
0&0&0&0\\
\Delta^*&0&t_\perp&0\\
0&0&0&0\\
\end{array}\right),
\\
&\Sigma_\mathbf{k}=-\frac{i}{2}\left(\begin{array}{cccc}
\Gamma^{11}_\mathbf{k}&\Gamma^{12}_\mathbf{k}&0&0\\
\Gamma^{21}_\mathbf{k} &\Gamma^{22}_\mathbf{k}&0&0\\
0&0&\Gamma^{11}_{\bar{\mathbf{k}}}&\Gamma^{12}_{\bar{\mathbf{k}}} \\
0&0 &\Gamma^{21}_{\bar{\mathbf{k}}}&\Gamma^{22}_{\bar{\mathbf{k}}}\\
\end{array}\right).
\end{split}
\end{equation}
For simplicity, we further ignore the momentum dependence in $\Gamma^0_\mathbf{k}$ and $\Gamma^{11}_\mathbf{k}$, assume $\Gamma^0_\mathbf{k}\equiv\Gamma^0$, $\Gamma^{11}_\mathbf{k}\equiv\Gamma^{11}$, $\Gamma^{12/21}_{\mathbf{k}}=\Gamma^{12/21}(\cos\mathbf{k}_x-\cos\mathbf{k}_y)$, and $\Gamma^{22}_{\mathbf{k}}=\Gamma^{22}(\cos\mathbf{k}_x-\cos\mathbf{k}_y)^2$, and set $\Gamma^{12}=\Gamma^{21}=r\Gamma^{22}$ and $\Gamma^{11}=r^2\Gamma^{22}$. Here $r$ is the ratio of the tunneling $t^a$ into $d_{z^2}$ and $d_{x^2-y^2}$ orbitals and $(\cos\mathbf{k}_x-\cos\mathbf{k}_y)$ accounts for the effect of the sign change of the $d_{x^2-y^2}$ wave function along $x$ and $y$ directions on $t^2_\mathbf{k}$ assuming the lead electrons have isotropic wave functions. We also require that $\Delta$, $\Gamma^0$, $\Gamma^{22}$, and  $r$ are real numbers.

\begin{figure}[t]
\centering
\includegraphics[width=\linewidth]{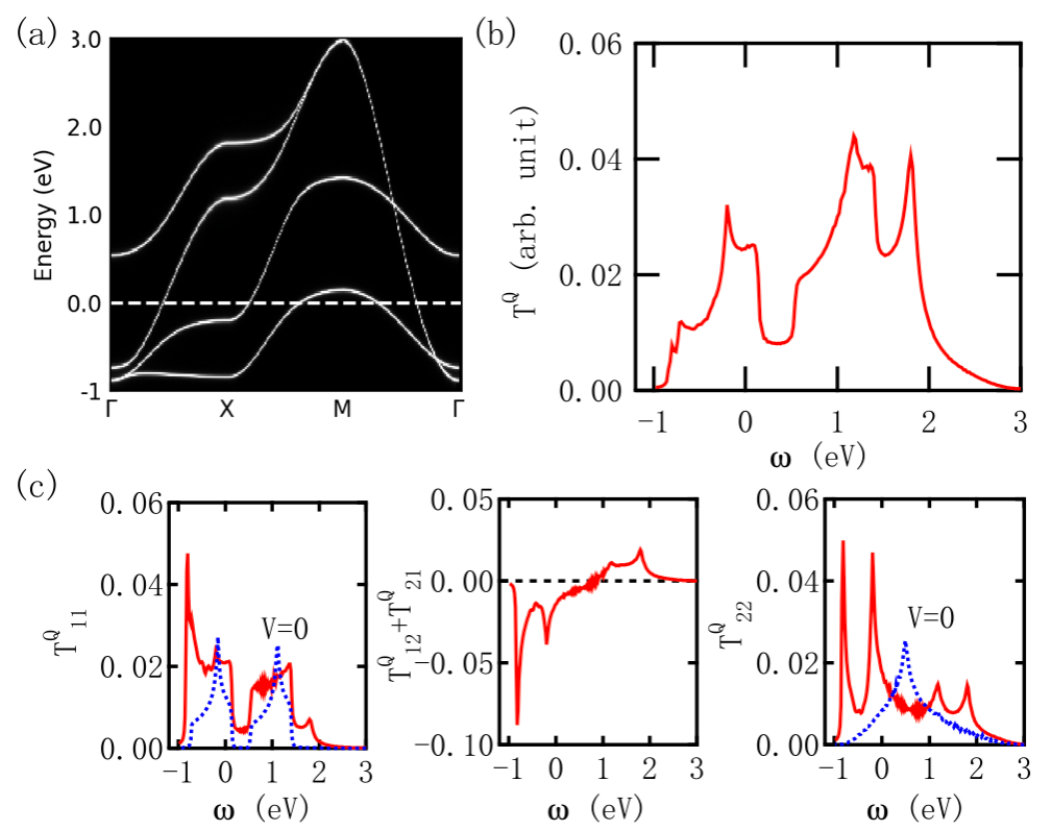}
\caption{(a) Spectral function calculated based on the tight-binding parameters and $\Gamma^0=\Gamma^{aa'}=0.01$. (b) The corresponding quasiparticle tunneling spectra showing multiple peaks. (c) Three components of the quasiparticle tunneling coefficient, showing a significant contribution from the off-diagonal term. The dotted lines show $T^Q_{11}$ and $T^Q_{22}$ at $V=0$ for comparison.}
\label{fig2}
\end{figure}

To verify the above formalism, we first show in Fig. \ref{fig2}(a) the calculated spectral function $A_\mathbf{k}(\omega)=-\pi^{-1}{\rm Im}\sum_{as} G^R_{1a\mathbf{k}s,1a\mathbf{k}s}(\omega)$ along a high symmetry path for $\Gamma^0=\Gamma^{22}=0.01$, $r=1$, and $\Delta=0$ using the tight-binding parameters. Direct comparisons with previous work confirm our derived band structures, which contains both bonding (around the Fermi energy) and antibonding bands of $d_{z^2}$ orbitals and several flat pieces induced by their hybridization with the broader $d_{x^2-y^2}$ bands \cite{DXYao2023}. Substituting the derived Green's functions in Eq. (\ref{TQTA}) gives the  quasiparticle tunneling coefficient $T^Q(\omega)$. Its numerical results are plotted in Fig. \ref{fig2}(b) and exhibit multiple pronounced peaks corresponding to the flat pieces of the bands, which reflect the interplay between the van Hove singularities and the hybridization between the $d_{z^2}$ and $d_{x^2-y^2}$ bands. For clarity, we also plot in Fig. 2(c) its three components, $T^Q_{11}$, $T^Q_{12}+T^Q_{21}$, and $T^Q_{22}$, where $T^Q_{aa'}=\sum_{a''\mathbf{k}s}\Gamma^0_{\mathbf{k}}\Gamma^{aa'}_{\mathbf{k}}G^R_{1a'\mathbf{k}s,1a''\mathbf{k}s}G^A_{1a''\mathbf{k}s,1a\mathbf{k}s}$. There exists a substantial contribution from the inteference term, $\mathcal{T}^Q_{12}+\mathcal{T}^Q_{21}$, in particular at negative energies where a sharp peak is almost completely cancelled out with those in $T^Q_{11}$ and $T^Q_{22}$. This corresponds to the flat band close to -1, which will be renormalized towards the Fermi energy once correlation effects are correctly taken into account. The valley between 0 and 0.5 only appears in $T^Q_{11}$, corresponding to the bonding-antibonding splitting region of the $d_{z^2}$ bands, below which the low-energy physics is governed by the bonding orbitals. For comparison, the quasiparticle tunneling coefficients without hybridization ($V=0$) are also shown in Fig. \ref{fig2}(c).

\begin{figure}[t]
\centering
\includegraphics[width=\linewidth]{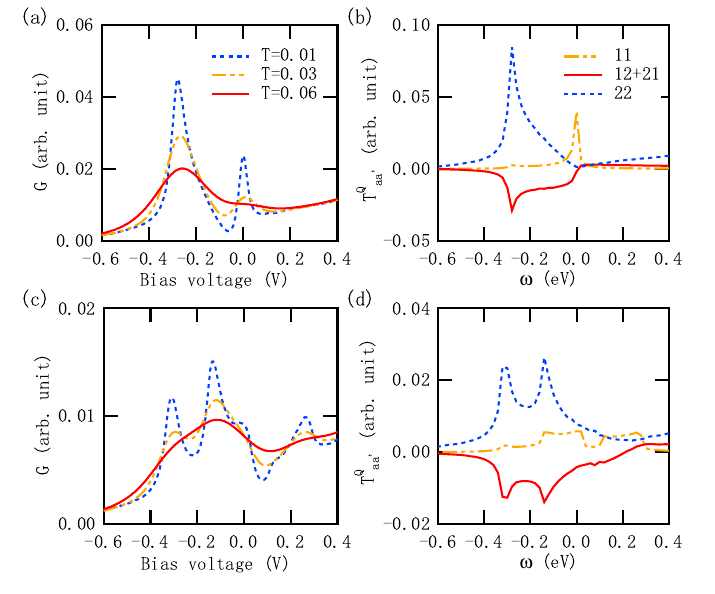}
\caption{(a) Differential conductance $G=dI(V_b)/dV_b$ from quasiparticle tunneling for the tight-binding parameters with $\eta_t=0$, $\eta_V=0.5$, $\Gamma^0=\Gamma^{22}=0.01$, and $r=0.2$ for the temperature $T=0.01$, 0.03, 0.06. (b) Three components of the corresponding quasiparticle tunneling coefficient, $T^Q_{11}$, $T^Q_{12}+T^Q_{21}$, and $T^Q_{22}$. (c) Differential conductance $G=dI/dV$ for the tight-binding parameters with $\eta_t=0.2$, $\eta_V=0.5$, $\Gamma^0=\Gamma^{22}=0.01$, and $r=0.3$ for $T=0.01$, 0.03, 0.06, showing additional peaks due to the bonding-antibonding splitting at low temperature but a broad asymmetric line shape at high temperature. (d) Three components of the quasiparticle tunneling coefficient, $T^Q_{11}$, $T^Q_{12}+T^Q_{21}$, and $T^Q_{22}$ for (c). For simplicity, we have ignored the temperature dependence of all parameters.}
\label{fig3}
\end{figure}

To have a better view of the interference effect in the low-energy region, we plot in Fig. \ref{fig3} the quasiparticle tunneling coefficient in a smaller energy window around the Fermi energy and take into consideration the renormalization effect by tuning $\epsilon^1_\mathbf{k}\rightarrow \eta_t \epsilon^1_\mathbf{k}$, $t_\perp\rightarrow \eta_t t_\perp$, and $V\rightarrow \eta_V V$, where $\eta_t$ and $\eta_V$ may be regarded to arise from the Gutzwiller projection for the nearly half-filled $d_{z^2}$ orbitals in La$_3$Ni$_2$O$_7$ but are taken here as free parameters. Figure 3(a) shows the temperature evolution of the differential conductance $G(V_b)=dI(V_b)/dV_b$ as a function of the bias voltage for $\eta_t=0$, $\eta_V=0.5$, $\Gamma^0=\Gamma^{22}=0.01$, and $r=0.2$. We have deliberately chosen the extreme case with $\eta_t=0$ to reflect a strongly renormalized flat $d_{z^2}$ quasiparticle bands, so that we may ignore other complications as discussed in Fig. \ref{fig2} and focus only on the hybridization effect in the quasiparticle tunneling. The ratio $r$ of the tunneling parameters into $d_{z^2}$ and $d_{x^2-y^2}$ orbitals should be correspondingly renormalized and therefore assigned a small number, which helps to suppress the $d_{z^2}$ quasiparticle weight ignored in the mean-field approximation. In the extreme case where $d_{z^2}$ electrons are fully localized, possibly at ambient pressure, we should have $\eta_t=\eta_V=r=0$ and the electrons can only tunnel into the $d_{x^2-y^2}$ orbital, giving rise to a broad background.

\begin{figure}[t]
\centering
\includegraphics[width=0.98\linewidth]{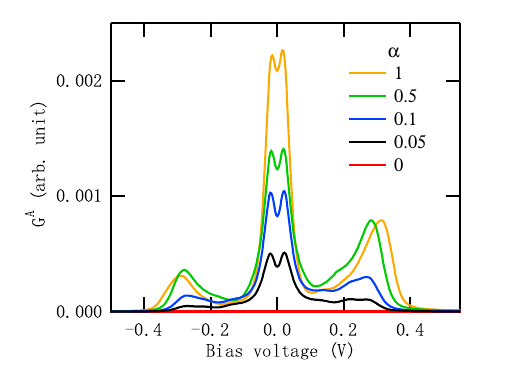}
\caption{(a) Different conductance contributed by the Andreev reflection coefficient using the tight-binding parameters with $\eta_t=0.2$, $\eta_V=0.5$, $\Gamma^0=\Gamma^{22}=0.01$, $r=0.3$, and the pairing order parameter $\Delta=0.02$ for the renormalized $t_\perp\rightarrow \alpha\eta_t t_\perp$ with varying $\alpha=1$, 0.5, 0.1, 0.05, 0 at $T=0.01$.}
\label{fig4}
\end{figure}

As is seen clearly in Fig. \ref{fig3}(a), the differential conductance $G(V_b)$ develops an asymmetric feature below the Fermi energy, which follows roughly a Fano-like line shape on top of the broad background from the $d_{x^2-y^2}$ metallic bands without hybridization. The suppression below $V=0$ reflects the destructive interference from the off-diagonal term $T^Q_{12}+T^Q_{21}$, which is not present without hybridization. This is better seen in Fig. \ref{fig3}(b), where we plot the three components of the quasiparticle tunneling coefficient $T^Q(\omega)$. The off-diagonal term remains almost unchanged around $\omega=-0.2$, which cancels out the diagonal contribution from $T^Q_{11}(\omega)$ and $T^Q_{22}(\omega)$ and yields the conductance minimum at negative $V$ slightly below zero bias. At low temperature, this suppression of $G(V_b)$ is interrupted by an unexpected peak at $V=0$. This peak only appears in $T^Q_{11}$  in Fig. \ref{fig3}(b) and is therefore a property solely of $d_{z^2}$. We attribute it to the nodal hybridization along the zone diagonal ($k_x=k_y$) direction, which leaves a finite contribution of the $d_{z^2}$ flat bands on the Fermi energy. In any case, the Fano interference effect is clearly identified to arise from the hybridization between the strongly renormalized flat $d_{z^2}$ quasiparticle bands and the broad metallic $d_{x^2-y^2}$ bands.

However, due to the complication of realistic electronic structures, it is not clear if the Fano effect can be easily resolved in experiment. To see this, we compare in Fig. \ref{fig3}(c) the differential conductance for a finite $\eta_t=0.2$, which corresponds to narrow but dispersive $d_{z^2}$ quasiparticle bands. Other parameters remain unchanged except for a slightly larger $r=0.3$. At low temperatures, the $G(V_b)$ curves become very different, where the single peak near -0.3 V in Fig. \ref{fig3}(a) splits into two, while the peak on the Fermi energy is suppressed. These changes arise from the finite dispersion and bonding-antibonding splitting of $d_{z^2}$ quasiparticle bands for $\eta_t=0.2$. As a result, there appear many additional peaks and the Fano feature is no longer discernible. However, at higher temperature $T=0.03$, all peaks get broadened and an overall asymmetric line shape can still be revealed on the broad background, reflecting a signature of the hybridization-induced interference. For completeness, Fig. \ref{fig3}(d) compares the three components of the corresponding quasiparticle tunneling coefficient. We see an equally large but negative off-diagonal contribution for the destructive interference. On the other hand, this off-diagonal term may be overwhelmed if the coupling of the lead with the $d_{z^2}$ orbital is too strong (large $r$). This is possible considering the $d_{z^2}$ quasiparticle wave function, though renormalized with a small spectral weight, is perpendicular to the NiO plane and may have a larger overlap with the lead electrons than that of the $d_{x^2-y^2}$ orbital. Then the differential conductance around zero bias would be governed by the features of the $d_{z^2}$ spectra as shown in $T^Q_{11}(\omega)$. We leave for future experiment to see if the hybridization-induced Fano interference effect may be resolved in real measurements of pressurized La$_3$Ni$_2$O$_7$, which may depend on the specific parameter range in experiment. Similar Fano phenomena have been predicted and observed previously in heavy-fermion materials \cite{Yang2019PRB}.

Now we turn to the superconducting state and study the Andreev reflection for interlayer pairing. Figure \ref{fig4} shows the typical differential conductance $G^A(V_b)$ contributed solely by the Andreev reflection coefficient $T^A(\omega)$ assuming $\Delta=0.02$ for the interlayer pairing. To study how it evolves with interlayer hopping, we further reduce $\eta_t t_\perp\rightarrow \alpha\eta_t t_\perp$ by multiplying an additional factor $\alpha=0.5$, 0.1, 0.05, and 0. Other parameters are the same as in Fig. \ref{fig3}(d). For finite $\alpha$, we see clearly the Andreev reflection peak around the zero bias. By contrast, one expects a V- or U-shape gap in the quasiparticle tunneling contribution. However, as $\alpha$ decreases, the height of the peak is gradually reduced and eventually diminishes for $\alpha=0$. Thus, the Andreev reflection is completely suppressed without interlayer hopping, which reveals a distinct feature compared to that of intralayer pairing superconductivity. The suppression is in some sense analogous to the junction of a fully polaized ferromagnet and a spin-singlet superconductor. It is easy to understand because Andreev reflection only occurs when incident electrons can form Cooper pairs in the superconductor and be reflected as holes, but this process is not possible for interlayer pairing superconductivity if the incident electrons can only tunnel into the top layer with $t_\perp=0$. A finite $t_\perp$ not only induces a higher-order intralayer pairing component but also allows for the tunneling into the bottom layer, thus contributing a finite Andreev reflection coefficient. For small $t_\perp$ that is greatly renormalized by strong electronic correlations and proportional to the self-doped hole density in nearly half-filled $d_{z^2}$ orbitals, our results indicate a much weaker Andreev reflection that are tuned systematically with pressure or $d_{z^2}$ hole doping, different from the usual intralayer pairing that has been extensively studied in previous literatures \cite{Deutscher2005RMP}. This is a fundamental difference that may help distinguish interlayer and intralayer pairing scenarios. However, one should be careful to exclude other factors such as the junction transparency that may also suppress the Andreev reflection.

It should be noted that we do not attempt to make any quantitative predictions on the experimental observations, since the latter may sensitively depend on the details of the setup such as the quality of the contact and the surface, as well as the parameter region of the compounds, let alone the difficulty in high pressure measurements. The mean-field approximation itself also cannot fully capture the strongly-correlated nature of the system. Neither do we expect to exclusively determine the pairing symmetry as that may require very hard work possibly based on phase sensitive measurement. We only want to point out the possible occurrence of the Fano interference effect in the presence of the hybridization and the suppression of the Andreev reflection in the absence of the interlayer tunneling, which are general properties of the interlayer pairing superconductivity in the $t$-$V$-$J$ model. These are qualitative conclusions that are independent of many details and hence may also be present in the trilayer La$_4$Ni$_3$O$_{10}$ system under the $t$-$V$-$J$ model \cite{Qin2024InnMat}. These features might be difficult to observe due to material complications considering the current experimental controversy. Nevertheless, one may still expect to get some useful information on the basic physics of nickelate high-temperature superconductors under pressure.

This work was supported by the National Natural Science Foundation of China (Grants No. 12474136), the Strategic Priority Research Program of the Chinese Academy of Sciences (Grant No. XDB33010100), and the National Key Research and Development Program of China (Grant No. 2022YFA1402203).

\end{document}